\documentclass[12pt]{article}
\usepackage{feynmf}
 
\newcommand{\be}{\begin{equation}}
\newcommand{\ee}{\end{equation}}
\newcommand{\bea}{\begin{eqnarray}}
\newcommand{\eea}{\end{eqnarray}}
\newcommand{\beann}{\begin{eqnarray*}}
\newcommand{\eeann}{\end{eqnarray*}}
\newcommand{\beasn}{\begin{sneqnarray}}
\newcommand{\eeasn}{\end{sneqnarray}}

\catcode`@=11
\def\dif{{\rm d}}
\def\deriv{\@ifnextchar[{\@deriv}{\@deriv[]}}
   \def\@deriv[#1]#2#3{\mathchoice%
{{\dif^{#1}#2\over\dif{#3}^{#1}}}{{\dif^{#1}#2/\dif{#3}^{#1}}}%
{{\dif^{#1}#2\over\dif{#3}^{#1}}}{{\dif^{#1}#2/\dif{#3}^{#1}}}}

\def\presup#1{{}^{#1}\kern-.15em\relax}      
\def\presub#1{{}_{#1}\kern-.12em\relax}      

\def\bfnabla{\mbox{\boldmath $\nabla$}}
\def\bfpi{\mbox{\boldmath $\pi$}}

\def\secteqno{\@addtoreset{equation}{section}%
\def\theequation{\thesection.\arabic{equation}}}
\def\endsecteqno{\def\theequation{\@ifundefined{chapter}%
{\arabic{equation}}{\thechapter.\arabic{equation}}}}
\newcounter{subequation}
\def\thesubequation{\alph{subequation}}
\def\sneqnarray{\stepcounter{equation}\let\@currentlabel=\theequation
\setcounter{subequation}{1}
\def\@eqnnum{{\rm (\theequation\thesubequation)}}
\global\@eqcnt\z@\tabskip\@centering\let\\=\@eqncr\let\@@eqncr=\@@sneqncr
$$\halign to \displaywidth\bgroup\@eqnsel\hskip\@centering
 $\displaystyle\tabskip\z@{##}$&\global\@eqcnt\@ne
 \hskip 2\arraycolsep \hfil${##}$\hfil
 &\global\@eqcnt\tw@ \hskip 2\arraycolsep $\displaystyle\tabskip\z@{##}$\hfil
  \tabskip\@centering&\llap{##}\tabskip\z@\cr}
\def\endsneqnarray{\@@sneqncr\egroup $$\global\@ignoretrue}
\def\@@sneqncr{\let\@tempa\relax
   \ifcase\@eqcnt \def\@tempa{& & &}\or \def\@tempa{& &}
   \else \def\@tempa{&}\fi
     \@tempa \if@eqnsw\@eqnnum\stepcounter{subequation}\fi
     \global\@eqnswtrue\global\@eqcnt\z@\cr}
\def\nobiblabels{\def\@lbibitem[##1]##2{\@bibitem{##2}}}
\catcode`@=12
\newcommand{\Appendix}[1]%
    {\renewcommand{\thesection}{Appendix~\Alph{section}}%
     \section{#1}%
     \renewcommand{\thesection}{\Alph{section}} }

\newcommand{\nn}{\nonumber}

\secteqno
\begin{document}

\def\s{\sigma}
\def\g{\gamma}
\def\m{\mu}
\def\n{\nu}
\def\d{\delta}
\def\als{\alpha_{s}}


\title{{\bf Effective Field Theory Approach to Pionium}}

\author{{\sc D.\,Eiras} \, and \,
        {\sc J.\,Soto}\\
        \small{\it{Departament d'Estructura i Constituents
               de la Mat{\`e}ria}}\\
        \small{\it{and Institut de F{\'\i}sica d'Altes Energies.}}\\
        \small{\it{Universitat de Barcelona, Diagonal, 647}}\\
        \small{\it{E-08028 Barcelona, Catalonia, Spain.}}\\
        {\it e-mails:} \small{dolors@ecm.ub.es, soto@ecm.ub.es} }

\date{\today}

\maketitle

\thispagestyle{empty}

\begin{abstract}

The various dynamical scales below the pion mass
 involved in $\pi^{+}$ $\pi^{-}$ atoms 
are
 sequentially integrated out using non-relativistic
 effective 
 field theory
techniques. This allows us to systematically organise the corrections to the energy
levels and decay width. We present our results in terms of a single 
unknown constant which may be obtained by matching to the Chiral Lagrangian 
with electromagnetic interactions at two loops.

\end{abstract}

\medskip

Keywords: Effective Field Theories, Hadronic Atoms, NRQED, Chiral Perturbation Theory.

PACS: 30.10.Gv, 11.10.St, 12.39.Fe, 13.75b 

\vfill
\vbox{
\hfill May 1999\null\par
\hfill UB-ECM-PF 99/10}\null\par

\clearpage



\section{Introduction}
\indent

\bigskip

Hadronic atoms have attracted much interest since long \cite{Deser}. Typically there is 
an interesting interplay between strong and electromagnetic interactions. Whereas the 
latter are the responsible of the bound state formation, the former produce their decay. 
Although the treatment of electromagnetic interactions is based on solid theoretical 
grounds, this is not so for the strong interactions. Traditionally, the latter are modeled 
by various types of short 
range potentials \cite{Rasche}. Although this is usually enough to fit the available data,
 it would be 
desirable to have a more direct connection with what is believed to be the fundamental 
theory of strong interactions, namely QCD. This is becoming even more urgent since the current 
DIRAC experiment at CERN \cite{Dirac}, which plans to measure the pionium decay width at $10\%$
accuracy \cite{Nemenov}, is meant to extract the pure hadronic pion-pion 
scattering lengths, which may, in principle, be obtained from QCD.

\medskip
It has become apparent during the last decade, that the most fruitful way to approach 
low energy strong interaction physics from QCD is not by direct calculations from this 
theory but going through intermediate effective field theories (EFT), which are equivalent 
to QCD in a particular range of energies. For instance the Chiral Lagrangian \cite{GL} 
is an EFT for pions, 
which is equivalent to QCD for energies below the rho mass. The EFTs typically depend 
on various unknown constants, which in principle may be obtained from the fundamental theory.    
In practise, this may sometimes be achieved, like for instance in the case of Non-Relativistic QED
(NRQED) \cite{Lepage}
 where the constants can be determined order by order 
in $\alpha$, but
 many times is beyond our current technical abilities, like in the case of the Chiral 
Lagrangian, which would require large lattice simulations with light dynamical quarks or 
yet-to-be-discovered alternative non-perturbative techniques with a good control on the 
mechanism of chiral symmetry breaking. In any case, if the number 
of constants is small enough, they can be phenomenologically obtained from available data and 
used later on to predict new results, as it is the case of the Chiral Lagrangian.

\medskip

Pionium is a $\pi^{+}$ $\pi^{-}$ electromagnetic bound state of binding energy $\sim 2 keV$
 which decays strongly, basically to two $\pi^{0}$, with a width $\Gamma \sim .6 eV$ \cite{pionium}.
 Clearly 
a QCD based analysis of this system should better start with the Chiral Lagrangian. However, 
the Chiral Lagrangian is a relativistic (manifestly Lorentz invariant) theory where electromagnetic
 bound state problems are difficult to handle (see \cite{Sazdjian, Akaki} for direct approaches).
 Moreover,
both the binding 
energy and the decay width are much smaller than the pion mass $\sim 140 MeV$, which suggests that a
 non-relativistic approach should be appropriated. 

\medskip

It is the aim of this work to present a non-relativistic approach to pionium based on a series of EFTs
which are obtained from the Chiral Lagrangian coupled to electromagnetism after sequentially integrating 
out the various physical 
scales of the system until we reach the scale of the binding energy
 $\sim m\alpha^2/4$. The first scale 
to be integrated out 
is the pion mass $m$. This produces a local non-relativistic EFT for pion pairs near threshold, 
much in the same way as NRQED is obtained from QED \cite{Lepage,NRQED, ManoharMatch}. The next relevant scale in 
the problem is the 
mass difference between charged and neutral pions $\Delta m \sim 5 MeV$. Integrating out this scale produces    
a local EFT with only charged pion fields in it. The next relevant scale is the typical relative 
momentum of pions in the bound state $m\alpha/2 \sim 0.5 MeV$ (soft). Integrating out this scale is, at lower orders
of $\alpha$, equivalent to calculating the electromagnetic potential between the two charged pions. The 
calculations in the latter EFT reduce to quantum mechanical ones. The main advantage of this 
approach is that
there are well defined counting rules at any stage of the calculation, so that the size of any 
neglected term
is easy to estimate. This is particularly important in order to extract more accurate values for
 the parameters of the Chiral Lagrangian  from the improved measurement
 of the pionium decay width in the DIRAC experiment 
\cite{Dirac}.

\medskip  

We distribute
the paper as follows.
 In Section 2 
 we present the most general non-relativistic effective field theory for pion pairs near threshold. The
constrains due to Lorentz invariance are implemented and the lagrangian is
reduced to its minimal form by local field redefinitions.
 In Section 3
the neutral pions are integrated out which gives rise to a non-relativistic theory of
charged pions interacting with the electromagnetic field. 
In Section 4 we integrate out soft photons, which produce the electromagnetic potentials
 between the charged
 pions. In Section 5 we present the calculation of
the bound state energies and decay widths.
Section 6 is devoted to the discussion of our results.
In Appendix A we discuss the realisation of Lorentz symmetry in non-relativistic theories. 
In Appendix B we display the local field redefinitions and the various 
 reshuffling of constants carried out  along the paper .
In Appendix C we present a new way to 
regulate the Coulomb propagator in $D$ space dimensions.

\newpage
\section{Non-relativistic lagrangian for pion pairs near 
$\quad\quad$ threshold}
\indent

\bigskip

At relative momentum much smaller than the pion mass a non-relativistic description of
pion pairs should be appropriated. In order to implement it, we shall write down a lagrangian 
organised in powers of
$1/m$ in which any scale smaller than $m$ is treated perturbatively. For the problem
at hand the next relevant energy and momentum scales are $\Delta m$ and 
 $\sqrt{m  \Delta m}$, its associated momentum, respectively. These scales are to be used 
to estimate the (maximum) size of each term.
\medskip
 
The symmetries (exact and approximate) of the fundamental theory, namely the
Chiral Lagrangian, must be incorporated. Let us consider first the internal
symmetries. The Chiral Lagrangian is approximately invariant under (non-linear) chiral
transformations, which are explicitely broken by the pion mass terms. Since the pion
mass is a large parameter in the non-relativistic lagrangian, no algebraic constraints from chiral
symmetry are expected to survive. All information about chiral symmetry will be
hidden in the parameters of the lagrangian. The only remaining approximate internal symmetry
will be isospin, which is explicitely broken by $m_u\not=m_d$ and the e.m. interactions
both at the quark and at the Chiral Lagrangian level. The size of the explicit breaking
may be estimated from $m_{\pi_{+}}-m_{\pi_0}\sim 5 MeV$ which is much smaller than the
pion mass. Hence isospin symmetry is a good (approximate) symmetry for the
non-relativistic lagrangian. In order to implement it we shall use the vector {\bf
$\bfpi$}

\be
\bfpi=({\pi_{+}+\pi_{-}\over \sqrt{2}}, {\pi_{-}-\pi_{+}\over \sqrt{2}i}, \pi_{0})
\ee
where $\pi_{+}$, $\pi_{-}$ and $\pi_{0}$ annihilate positive, negative and neutral
pions respectively.

\medskip

Concerning the space-time symmetries, Poincar\'e invariance (including the discrete
symmetries) must also be implemented in the non-relativistic lagrangian. The
translational and rotational part of the Poincar\'e group as well as the discrete
symmetries are implemented in the standard way. The Lorentz subgroup requires the
introduction of a non-linear realisation which 
 is equivalent to impose the so called reparametrisation
invariance \cite{LM}. This is discussed in Appendix A. The outcome is relatively
simple for spin zero fields. Consider a composite spin zero field made out of tensor
products of $n$ {\bf $\bfpi$} and $m$ {\bf $\bfpi^{\dagger}$}. Define $w=n-m$ the weight of
this field.  If $w\not= 0$, all derivatives acting on this field must be introduced
through the combination

\be
D=i\partial_0 -{1\over 2wm}\partial_{\mu}\partial^{\mu}
\ee   
 
If $w=0$, $\partial_{\mu}$ on this field can be introduced. The lagrangian must have
all the Lorentz indices contracted in a formally Lorentz invariant way and $D$ must be
considered Lorentz invariant itself.
  
Having in mind the rules above, consider first the limit of exact isospin symmetry.
 We have 

\bea
{\rm L}=&& {\rm L}_2 +{\rm L}_4 \nn\\
{\rm L}_2=&& {\rm L}_2^{(0)}+{\rm L}_2^{(1)}+ ...\nn\\
{\rm L}_4=&& {\rm L}_2^{(1/2)}+{\rm L}_2^{(3/2)}+ ...\nn\\
 {\rm L}_2^{(0)}=&& \bfpi^{\dagger} D \bfpi\nn\\
 {\rm L}_2^{(1)}=&& \bfpi^{\dagger}
 A_0 D^2\bfpi \nn\\
 {\rm L}_4^{(1/2)}=&& B_1(\bfpi^{\dagger}  \bfpi)^2 +
B_2 (\bfpi  \bfpi)(\bfpi^{\dagger}  \bfpi^{\dagger})\nn\\
 {\rm L}_4^{(3/2)}=&&
A_1 (\bfpi D \bfpi)(\bfpi^{\dagger}  \bfpi^{\dagger})+ h.c.\nn\\
&&
 A_2(\bfpi^{\dagger} D \bfpi)(\bfpi^{\dagger}  \bfpi) + h.c.\label{iso}\\
&&A_3(\bfpi^{\dagger}  \bfpi^{\dagger}) D(\bfpi  \bfpi)\nn\\
&&A_4 \partial_{\mu}(\bfpi^{\dagger}  \bfpi)\partial^{\mu}(\bfpi^{\dagger}  {\bf
\bfpi})\nn\\
&&A_5 ({\bfpi^{\dagger}}^{i}  {\bfpi^{\dagger}}^{j}) D (\bfpi^{i}  \bfpi^{j})\nn
\eea

Consider next the isospin breaking terms. These may be due to e.m. interactions at the
quark level, e.m. interactions in the relativistic Chiral Lagrangian and $m_u\not=m_d$.
The electromagnetic interactions at quark level have an isospin invariant piece which
is absorbed in the constants (\ref{iso}). The e.m. isospin breaking pieces, both at
quark level and in the Chiral Lagrangian, are proportional to
$T^3$, and so is the isospin breaking piece due to $m_u\not=m_d$. Hence, in order to
incorporate isospin breaking effects in the non-relativistic lagrangian, it is enough
to construct further invariants with the vectors ${\bf Q}\sim (0,0,e)$ and ${\bf M}\sim
(0,0, m_u-m_d)$, taking into account that {\bf Q} must always appear in pairs. Although 
there is no extra 
difficulty in taking ${\bf M}$ into account, we shall ignore it here since,
due to charge conjugation, it appears quadratically and turns out 
to be very small  \cite{isob}. Then the e.m. isospin breaking terms read

\bea
\Delta{\rm L}=&&\Delta{\rm L}_2 +\Delta{\rm L}_4\nn\\
\Delta{\rm L}_2=&& \Delta{\rm L}_2^{(0)} +\Delta{\rm L}_2^{(1)}\nn\\
\Delta{\rm L}_4=&& \Delta{\rm L}_2^{(3/2)} \nn\\
\Delta{\rm L}_2^{(0)}=&& \d_1 ({\bf \bfpi^{\dagger}  Q})({\bf Q}\bfpi)\nn\\
\Delta{\rm L}_2^{(1)}=&& \d_2 ({\bfpi^{\dagger}} {\bf Q})D ({\bf Q} \bfpi )\label{isob}\\
\Delta{\rm L}_4^{(3/2)}=&&C_1 (\bfpi  {\bf Q})(\bfpi  {\bf Q})
(\bfpi^{\dagger}  \bfpi^{\dagger})
+ h.c.\nn\\
&&+ C_2 (\bfpi  {\bf Q})(\bfpi^{\dagger}  {\bf Q})
(\bfpi^{\dagger}  \bfpi)\nn\\
&&+ C_3 ((\bfpi^{\dagger} \times \bfpi)\cdot {\bf Q})^2\nn
\eea

Before going on, let us discuss the general structure of the constants $A_{i}$, $B_{i}$, 
$C_{i}$ and $\delta_{i}$ above. Let us call $Z$ to any such a constant and $z$  its dimension. Then 
the general form of $Z$ will be
\bea
Z=&&m^{z}\Bigl( a_{-1} + a_0 ({m\over 4\pi f})^2+ a_1 ({m\over 4\pi f})^4+ a_2 ({m\over 4\pi f})^6+...\nn\\
&& + b_1 \alpha + ...\\
&& + c_{1,1} \alpha  ({m\over 4\pi f})^2 + c_{1,2} \alpha  ({m\over 4\pi f})^4 + ....\Bigr)\nn
\eea
where  $f\sim 93 MeV$ is the pion decay constant.
The $a_{i}$, $i=-1,0,1,...$ stand for pure strong interaction contributions. It is interesting to 
notice that spontaneous chiral symmetry breaking implies $a_{-1}=0$ for $Z\not=A_0,\, \delta_{i}$. Indeed 
in the limit $f\rightarrow \infty $ 
(keeping  $m$ constant) the pions in the Chiral Lagrangian become free particles as far as 
 the strong interactions is concerned. Hence, in this limit any EFT
 derived from the Chiral Lagrangian must not contain strong interactions. Then the subscript $i=0,1,..$
coincides with the number of loops at which the term  $a_{i}$ receives contributions. We stopped at
 the number of loops which have been calculated so far \cite{2loop}. $b_{i}$, $i=1,2,,...$ stand for purely 
electromagnetic contributions and $c_{i,j}$, $i,j=1,2,,...$ for mixed electromagnetic and strong contributions. 
We stop here at the orders which compare to the two loop purely strong contribution.  $b_{1}$ may 
receive contributions from tree level annihilation graphs, $c_{1,1}$ from one loop graphs \cite{Knech} and 
$c_{1,2}$ from two loop graphs yet to be calculated. For this discussion to apply to the constants 
$C_{i}$ and $\delta_{i}$ of the isospin breaking terms ${\bf Q}$ must be counted as a dimension one object.
  
\medskip 

The lagrangian (\ref{iso}) and  (\ref{isob}) contains higher time derivative terms. One can get rid of these terms by
local field redefinitions. 
 We can
set $A_0=\d_2=A_1=A_2=0$ by using local field redefinitions which 
maintain Lorentz symmetry explicit.
However, the new lagrangian still contains time derivatives  beyond the expected $i\partial_0$. We
can also get rid of the extra time derivatives by using again local field redefinitions,
which cannot maintain Lorentz symmetry explicit anymore. The details of this are displayed in
Appendix B. We finally obtain the lagrangian in the so called minimal form

\bea
L=&&L_2+L_4\nn\\
L_2=&& {\bfpi^{\dagger}}^{j}\Big( (i\partial_0 +{\bfnabla^2\over 2m} + {\bfnabla^4\over
8m^3})\delta_{ij} \nn\\
&& +(1+{\bfnabla^2\over 2m^2})\Delta m {{\bf Q}^{i}{\bf Q}^{j}\over {\bf Q}^2}
\Big)
 \bfpi^{i} \nn\\
L_4=&& B_1(\bfpi^{\dagger}  \bfpi)^2 +
B_2 (\bfpi  \bfpi)(\bfpi^{\dagger}  \bfpi^{\dagger})\nn\\
&& +  D_1(\bfpi^{\dagger}{\bfnabla^2\over 2m}  \bfpi
+\bfpi{\bfnabla^2\over 2m}  \bfpi^{\dagger})
(\bfpi^{\dagger}  \bfpi) \label{minimal}\\
&&+D_2 \Bigl( (\bfpi{\bfnabla^2\over 2m}  \bfpi)\bfpi^{\dagger}  \bfpi^{\dagger}
+\bfpi  \bfpi(\bfpi^{\dagger}{\bfnabla^2\over 2m}  \bfpi^{\dagger})\Bigr)\nn\\  
&& + 2A_4 (\bfpi^{\dagger}  \bfpi)\partial^{i}\bfpi^{\dagger} \partial^{i}  {\bf
\bfpi}\nn\\
&&+C_1^{\prime} (\bfpi  {\bf Q})(\bfpi  {\bf Q})
(\bfpi^{\dagger}  \bfpi^{\dagger})+ h.c.\nn\\
&&+ C_2^{\prime} (\bfpi  {\bf Q})(\bfpi^{\dagger}  {\bf Q})(\bfpi^{\dagger}  \bfpi)\nn\\
&&+ C_3 ((\bfpi^{\dagger} \times \bfpi)\cdot {\bf Q})^2 \nn\\
&&+{A_3\over 2}(\bfpi^{\dagger}  \bfpi^{\dagger}){\bfnabla^2\over 2m} (\bfpi  \bfpi ) \nn\\
&&+{A_5\over 2}({\bfpi^{\dagger}}^{i}  {\bfpi^{\dagger}}^{j}){\bfnabla^2\over 2m} (\bfpi^{i}  \bfpi^{j} ) \nn
\eea
The new constants above are defined in formula (B.8) of Appendix B. Lorentz symmetry guarantees that the bilinear 
terms have the standard form including relativistic corrections. It also relates $A_3$ and $A_5$ in the two
last terms to the remaining constants (see Appendix B). Unfortunately, the latter relations have no practical 
consequences because the two last terms are proportional  
to the center of mass momentum and hence irrelevant to our problem.  The zero charge sector in terms of 
the pion field reads
\bea
L_2=&&\pi_{+}^{\dagger} (i\partial_0 +{\bfnabla^2\over 2m} + {\bfnabla^4\over
8m^3})\pi_{+} 
 +\pi^{\dagger}_{-} (i\partial_0 +{\bfnabla^2\over 2m} + {\bfnabla^4\over
8m^3})\pi_{-} \nn\\
&&\quad  +\pi^{\dagger}_{0} (i\partial_0 +\Delta m +{\bfnabla^2\over 2m} +
\Delta m {\bfnabla^2\over 2m^2}+
 {\bfnabla^4\over
8m^3})\pi_{0} \nn\\
L_4=&&  R_{00} \pi^{\dagger}_0   \pi^{\dagger}_0 \pi_0  \pi_0 +
 R_{cc} \pi^{\dagger}_{+}   \pi^{\dagger}_{-} \pi_{+}  \pi_{-}
+( R_{0c} \pi^{\dagger}_{0}   \pi^{\dagger}_{0} \pi_{+}  \pi_{-} + h.c. )\label{zero}\\
&& + S_{00}( \pi^{\dagger}_0   \pi^{\dagger}_0 \pi_0 \bfnabla^2 \pi_0 + h.c. )
+ S_{cc} \Big( \pi^{\dagger}_{+}   \pi^{\dagger}_{-}( \pi_{+}\bfnabla^2  \pi_{-}
+\pi_{-}\bfnabla^2  \pi_{+})+\; h.c.\Big)\nn\\ 
 && +
S_{0c} \Big( \pi^{\dagger}_{0}   \pi^{\dagger}_{0}( \pi_{+}\bfnabla^2  \pi_{-}
+\pi_{-}\bfnabla^2  \pi_{+}) +2\pi^{\dagger}_{+}   \pi^{\dagger}_{-} \pi_{0}\bfnabla^2  \pi_{0}
+ h.c. \Big)
\nn\\&&+\quad  P_{00} \pi^{\dagger}_0 \partial_{i}  \pi^{\dagger}_0 \pi_0 \partial_{i} \pi_0 +
 P_{cc}( \pi^{\dagger}_{+} \partial_{i}  \pi^{\dagger}_{-} \pi_{+} \partial_{i} \pi_{-}
+  \pi^{\dagger}_{-} \partial_{i}  \pi^{\dagger}_{+} \pi_{-} \partial_{i} \pi_{+}) \nn
\eea
The new constants above are defined in formula (B.9) of Appendix B.
Notice that since the origin of energies appears to be at the two charged pion
threshold,
the neutral pion shows a negative energy gap $-\Delta m < 0$. Notice also that the terms in the
bilinear neutral pion lagrangian can be combined into the standard form.
\be
 (i\partial_0 +\Delta m +{\bfnabla^2\over 2m} +
\Delta m {\bfnabla^2\over 2m^2}+
 {\bfnabla^4\over
8m^3}) \sim (i\partial_0 +\Delta m +{\bfnabla^2\over 2(m-\Delta m)} + {\bfnabla^4\over
8(m-\Delta m)^3})
\ee
Nevertheless, in order to keep the expansion systematic we shall not use the
expression above.

\medskip

The coupling to e.m. fields is done by promoting normal derivatives to covariant
derivatives. None of the possible non-minimal couplings contributes at the order we
are interested in and we will ignore them.

\medskip

Before closing this section let us remark that we have assumed that the lagrangian (\ref{iso}) and 
(\ref{isob}) is hermitian. This is correct at the order we are interesting in. However, in general the hermiticity
constraint must be relaxed. This is due to the fact that the $\pi^{+}$$\pi^{-}$ atom may decay into degrees of
 freedom which do not appear in the non relativistic lagrangian, for instance to hard photons or hard 
electron-positron pairs. The non-hermitian pieces would be obtained in the matching to the Chiral Lagrangian 
at the same time as the hermitian ones, as it happens in NRQED 
\cite{NRQED,ManoharMatch,LambPos}. 
\medskip


 \begin{figure}

 \begin{fmffile}{10deltam}
 \begin{center}{
 \begin{fmfgraph*}(90,90)
 \fmfleft{i1,i2}\fmfright{o1,o2}
 \fmf{fermion,label=$\pi^{+}$,label.side=left}{i1,v1}\fmf{fermion,label=$\pi^{-}$,label.side=right}{i2,v1} 
 \fmf{fermion,label=$\pi^{+}$,label.side=left}{v2,o1}\fmf{fermion,label=$\pi^{-}$,label.side=right}{v2,o2} 
 \fmf{fermion,left,tension=.2,width=thin,label=$\pi_0$,label.side=left}{v1,v2}
 \fmf{fermion,right,tension=.2,width=thin,label=$\pi_0$,label.side=right}{v1,v2}
 \end{fmfgraph*}
\hspace*{1cm} 
 \begin{fmfgraph*}(90,90)
 \fmfleft{i1,i2}\fmfright{o1,o2}
 \fmf{fermion,label=$\pi^{+}$,label.side=left}{i1,v1}\fmf{fermion,label=$\pi^{-}$,label.side=right}{i2,v1} 
 \fmf{fermion,label=$\pi^{+}$,label.side=left}{v2,o1}\fmf{fermion,label=$\pi^{-}$,label.side=right}{v2,o2} 
 \fmf{fermion,left,tension=.2,width=thin,label=$\pi^0$,label.side=left}{v1,v3}
 \fmf{fermion,right,tension=.2,width=thin,label=$\pi^0$,label.side=right}{v1,v3} 
 \fmf{fermion,left,tension=.2,width=thin,label=$\pi^0$,label.side=left}{v3,v2}
 \fmf{fermion,right,tension=.2,width=thin,label=$\pi^0$,label.side=right}{v3,v2}
 \end{fmfgraph*}
 \hspace*{1cm}
 \begin{fmfgraph*}(90,90)
 \fmfleft{i1,i2}\fmfright{o1,o2}
 \fmf{fermion,label=$\pi^{+}$,label.side=left}{i1,v1}\fmf{fermion,label=$\pi^{-}$,label.side=right}{i2,v1} 
 \fmf{fermion,label=$\pi^{+}$,label.side=left}{v2,o1}\fmf{fermion,label=$\pi^{-}$,label.side=right}{v2,o2} 
 \fmf{fermion,left,tension=.2,width=thin,label=$\pi^0$,label.side=left}{v1,v3}
 \fmf{fermion,right,tension=.2,width=thin,label=$\pi^0$,label.side=right}{v1,v3} 
 \fmf{fermion,left,tension=.2,width=thin,label=$\pi^0$,label.side=left}{v3,v4}
 \fmf{fermion,right,tension=.2,width=thin,label=$\pi^0$,label.side=right}{v3,v4} 
 \fmf{fermion,left,tension=.2,width=thin,label=$\pi^0$,label.side=left}{v4,v2}
 \fmf{fermion,right,tension=.2,width=thin,label=$\pi^0$,label.side=right}{v4,v2} 
 \end{fmfgraph*}
}\end{center}
 \vspace*{2cm}
\begin{center}{
 \begin{fmfgraph*}(90,90)
 \fmfleft{i1,i2}\fmfright{o1,o2}
 \fmf{fermion,label=$\pi^{+}$,label.side=left}{i1,v1}\fmf{fermion,label=$\pi^{-}$,label.side=right}{i2,v1} 
 \fmf{fermion,label=$\pi^{+}$,label.side=left}{v2,o1}\fmf{fermion,label=$\pi^{-}$,label.side=right}{v2,o2} 
 \fmf{fermion,left,tension=.2,width=thin,label=$\pi^0$,label.side=left}{v1,v2}
 \fmf{fermion,right,tension=.2,width=thin,label=$\pi^0$,label.side=right}{v1,v2}
 \fmfv{label=$S_{0c}$,decor.shape=hexacross,decor.size=60,label.side=right }{v2}
 \end{fmfgraph*}
\hspace*{1cm}
 \begin{fmfgraph*}(90,90)
 \fmfleft{i1,i2}\fmfright{o1,o2}\fmftop{t1}\fmfbottom{b1}
 \fmf{fermion,label=$\pi^{+}$,label.side=left}{i1,v1}\fmf{fermion,label=$\pi^{-}$,label.side=right}{i2,v1} 
 \fmf{fermion,label=$\pi^{+}$,label.side=left}{v2,o1}\fmf{fermion,label=$\pi^{-}$,label.side=right}{v2,o2} 
 \fmf{plain,left,tension=.2,width=thin,label=$\pi^0$,label.side=right}{v1,v2}
 \fmf{fermion,right,tension=.2,width=thin,label=$\pi^0$,label.side=right}{v1,v2}
 \fmf{phantom}{t1,v3}\fmf{phantom}{b1,v4}
 \fmf{phantom,left,tension=.2}{v3,v4,v3}
 \fmfv{decor.shape=circle,decor.filled=full,decor.size=60,label.side=down}{v3}
 \end{fmfgraph*}
\hspace*{1cm}
 \begin{fmfgraph*}(90,90)
 \fmfleft{i1,i2}\fmfright{o1,o2}
\fmftop{t1}\fmfbottom{b1}
 \fmf{fermion,label=$\pi^{+}$,label.side=left}{i1,v1}\fmf{fermion,label=$\pi^{-}$,label.side=right}{i2,v1} 
 \fmf{fermion,label=$\pi^{+}$,label.side=left}{v2,o1}\fmf{fermion,label=$\pi^{-}$,label.side=right}{v2,o2} 
 \fmf{plain,left,tension=.2,width=thin,label=$\pi^0$,label.side=right}{v1,v2}
 \fmf{fermion,right,tension=.2,width=thin,label=$\pi^0$,label.side=right}{v1,v2}
 \fmf{phantom}{t1,v3}
\fmf{phantom}{b1,v4}
 \fmf{phantom,left,tension=.2}{v3,v4,v3}
 \fmfv{decor.shape=triagram,decor.filled=full,decor.size=60,label.side=down}{v3}
 \end{fmfgraph*}
}\end{center}
 \end{fmffile}
\label{fig1}
\caption{Diagrams contributing to the matching between $L$ and $L^{\prime}$ up to corrections $O((\Delta m/m)^2)$.
The bullet and triangle insertions in the neutral pion propagator correspond to relativistic corrections due to 
$\bfnabla^4 / 8m^3$ and $\Delta m \bfnabla^2 / 2m^2$ respectively.}


\end{figure}
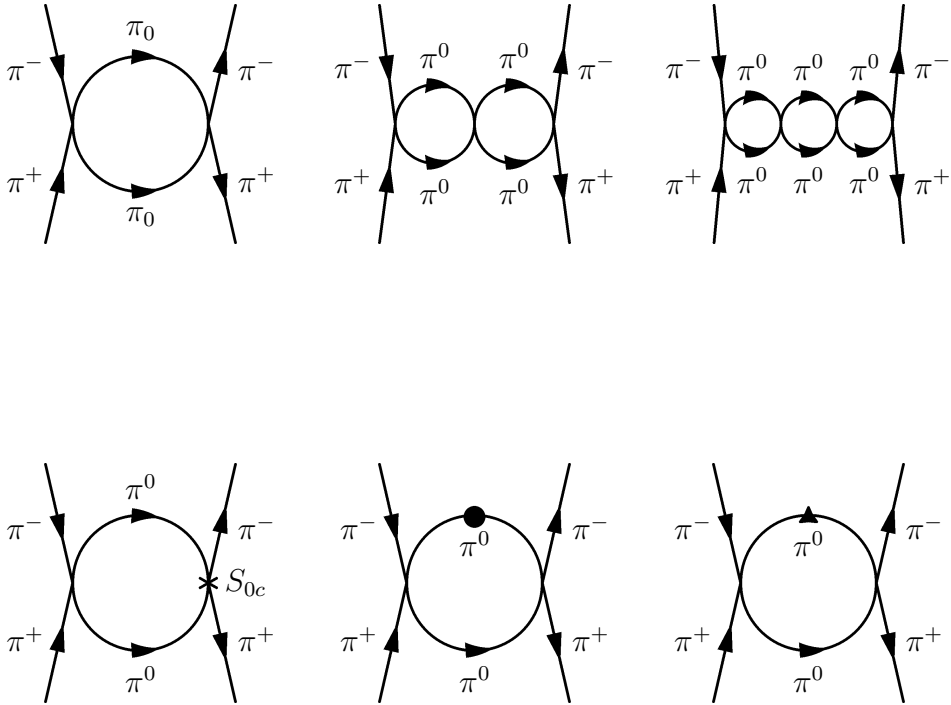

\section{Integrating out the scale $\Delta m$}
\indent
\bigskip

Since $\Delta m >> m\alpha^2/4$ it is appropiated to integrate out this scale before
tackling the e.m. bound state problem. This represents the main advantadge of our
 approach with respect to 
previous non-relativistic proposals \cite{Rasche, Ravndal, Holstein}. The integration of neutral 
pions can be easily achieved by
matching four point off-shell Green functions of the lagrangian above to a
non-relativistic lagrangian
where the neutral pions have been removed.
\newpage
\bea
L^{\prime}=&&  \pi_{+}^{\dagger} (iD_0 +{{\bf D}^2\over 2m} 
)\pi_{+} \nn\\
&& +\pi^{\dagger}_{-}(iD_0 +{{\bf D}^2\over 2m}
)\pi_{-} \label{charged}\\
&& +  R_{cc}^{\prime} \pi^{\dagger}_{+}   \pi^{\dagger}_{-} \pi_{+}  \pi_{-} +
 P^{\prime} \pi^{\dagger}_{+}   \pi^{\dagger}_{-} i\partial_0 \pi_{+}  \pi_{-} \nn \eea

 Since the $\pi^0$ energy gap is negative, the
integration will produce imaginary parts in $ R_{cc}^{\prime}$ and 
$ P^{\prime} $. By calculating the diagrams in Fig. 1 in dimensional regularisation (DR) we
obtain 

\bea
 R_{cc}^{\prime}=&& R_{cc}-\vert R_{0c}\vert^2 R_{00}\left(
{ms\over 2\pi}\right)^2\\
&& +i\vert R_{0c}\vert^2 {ms\over 2\pi}\Bigl( 1+ {5\over 8}{s^2\over m^2}
 -{3\over 4}{s^2\over m^2}
-
\left( {R_{00} ms\over 2\pi} \right)^2
-{2S_{0c}(R_{0c}+R_{0c}^{\ast}) s^2\over \vert R_{0c}\vert^2 } \Bigr) \nn\\
P^{\prime}= && i\vert R_{0c}\vert^2 {m^2\over 4\pi s} 
\eea
where $s=\sqrt{2m\Delta m}$. $ R_{cc}^{\prime}$ and $P^{\prime}$ contain the
 leading corrections in $\Delta m / m$ and $m\alpha^2 / 4\Delta m$ respectively.

\medskip

The electromagnetic contributions to $L^{\prime}$ coming from
the energy scale $\Delta m$ are neglegible, as well as the relativistic corrections $\sim \bfnabla^4/8m^3$ to the charge pions and the terms
 $P_{cc}$ and $S_{cc}$  in (2.7).


\section{Integrating out the scale $m\alpha$}
\indent

\bigskip
 The lagrangian in the previous section is almost identical to NRQED (for spin zero
particles) plus small local interactions. In refs. \cite{LambPos} it was shown that we 
can integrate out next dynamical scale, namely, $m\alpha /2$ in NRQED obtaining a further 
effective theory called potential NRQED
(pNRQED) which contains the usual potential terms and only the ultrasoft degrees of
freedom ($\sim m\alpha^2 / 4$) remain dynamical. We shall do the same here.
The (maximum) size of each term in (3.1) is obtained by assigning $m\alpha$ to any scale which is not
explicit.
 In fact, since we
are only interested in $O(\alpha)$ corrections, only the Coulomb potential seems to be
important, since the tranverse photons give rise to $O(\alpha^2)$ corrections. However,
as pointed out in ref. \cite{Labelle}, below the pion threshold there are further light
degrees of freedom apart from the photon. In particular, the electron mass $m_{e}\sim
m\alpha / 2$ and hence it must be integrated out here. This gives rise to a potential term
which is only $O(\alpha)$ suppressed with respect to the Coulomb one. By calculating the diagrams in Fig. 2 we obtain

\bea
L^{\prime\prime}=&&  \pi_{+}^{\dagger}({\bf x},t) (i\partial_0 +{\bfnabla^2\over 2m} )
\pi_{+}({\bf x},t) \nn\\ &&
 +\pi^{\dagger}_{-}({\bf x},t)(i\partial_0 +{\bfnabla^2\over 2m} )\pi_{-}({\bf x},t)
\nn\\ &&  +  R_{cc}^{\prime} (\pi^{\dagger}_{+}   \pi^{\dagger}_{-} \pi_{+}  \pi_{-})({\bf x},t)
+ P^{\prime} (\pi^{\dagger}_{+}   \pi^{\dagger}_{-})({\bf x},t) i\partial_0 (\pi_{+}  \pi_{-})({\bf x},t)
 \\
&& - \int d^3 {\bf y} (\pi_{+}^{\dagger}\pi_{+})({\bf x},t)\Bigr( V_0 (\vert {\bf x}-
{\bf y}\vert )+  V_1 (\vert {\bf x}-
{\bf y}\vert ) \Bigr)(\pi_{-}^{\dagger} \pi_{-})({\bf y},t)\nn\\
&& \nn\\ && \nn\\
&&  V_0 (\vert {\bf x}-
{\bf y}\vert )=-{\alpha\over \vert {\bf x}-
{\bf y}\vert } \quad , \quad  V_1 (\vert {\bf x}-
{\bf y}\vert )= \int {d^3 {\bf k} \over (2\pi)^3} V_{vpc} ({\bf k})e^{i({\bf x}-
{\bf y}){\bf k}}
\label{potential1}
\eea
where $ V_{vpc} ({\bf k})$ is given in formula (10) of ref. \cite{Labelle}. 
The lagrangian above contains no further degree of freedom than the non relativistic
charged pions and hence it is totally equivalent to standard quantum mechanics. We like better
to stay within the lagrangian formalism and use the $\pi^{-}$ $\pi^{+}$ wave function field 
$\phi({\bf x},{\bf X},t)$, where ${\bf x}$ and ${\bf X}$ are the relative and center of mass
 coordinates respectively, as introduced in \cite{LambPos}.     
\bea
L^{\prime\prime}=&&\phi^{\dagger}({\bf x},{\bf X},t)\Bigl(i\partial_0 +{\bfnabla^2\over m}
\nn\\ &&- V_0 (\vert {\bf x}\vert )-  V_1 (\vert {\bf x}\vert )
\label{potential2}\\ &&+ R_{cc}^{\prime}\delta({\bf x})+ P^{\prime}
\delta({\bf x})i\partial_0 \Bigr)\phi({\bf x},{\bf X},t)\nn
\eea
\bigskip 
The center of mass kinetic term has been dropped. 


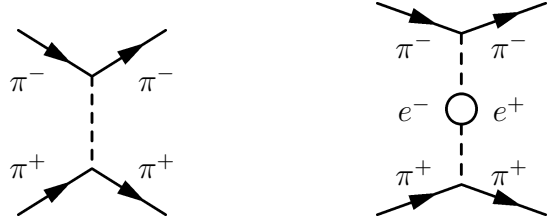
\begin{figure}
\begin{fmffile}{3fcoul}
 \begin{center}
 \begin{fmfgraph*}(70,70)
\fmfleft{i1,i2}\fmfright{o1,o2}
 \fmf{fermion,label=$\pi^+$,label.side=left}{i1,v1}
 \fmf{fermion,label=$\pi^-$}{i2,v2} 
\fmf{dashes}{v2,v1}
\fmf{fermion,label=$\pi^+$,label.side=left}{v1,o1} 
\fmf{fermion, label=$\pi^-$,label.side=right}{v2,o2}
 \end{fmfgraph*} 
 \hspace*{2cm}
 \begin{fmfgraph*}(80,80)
\fmfleft{i1,i2}\fmfright{o1,o2}
 \fmf{fermion,label=$\pi^+$,label.side=left}{i1,v1}
 \fmf{fermion,label=$\pi^-$}{i2,v2} 
\fmf{dashes}{v2,v3}
\fmf{dashes}{v4,v1}
\fmf{plain,left,label=$e^+$,label.side=left}{v3,v4}
\fmf{plain,right,label=$e^-$,label.side=right}{v3,v4}
\fmf{fermion,label=$\pi^+$,label.side=left}{v1,o1} 
\fmf{fermion, label=$\pi^-$,label.side=right}{v2,o2}
 \end{fmfgraph*} 
 \end{center}
\end{fmffile} 
\label{fig2}
\caption{Diagrams contributing to the matching between $L^{\prime}$ and $L^{\prime\prime}$ up to corrections 
$O(\alpha^2)$. Dashed lines are longitudinal photon propagators in the Coulomb gauge.}
\end{figure}

\section{Quantum mechanical calculation}
\indent
\bigskip

In order to calculate the corrections to the energy levels and decay width we shall
consider the propagator of (\ref{potential2}) and identify its pole. 
At the order we
are interested in only the diagrams in Fig. 3 contribute. 

\medskip

The diagrams in the first line of Fig. 3 correspond to first order perturbation theory and can be easily evaluated.  
They give rise to
 \bea
&\delta_{R_{cc}^{\prime}} E^{(1)}_n= -Re( R_{cc}^{\prime})
\vert \Psi_{n}({\bf 0})\vert^2   & \delta_{R_{cc}^{\prime}}
 \Gamma^{(1)}_n= 
2Im( R_{cc}^{\prime})\vert \Psi_{n}({\bf 0})\vert^2\\
&\delta_{P^{\prime}} E^{(1)}_n= 0\quad &  \delta_{P^{\prime}} \Gamma^{(1)}_n=
-Im(P^{\prime})\vert \Psi_{n}({\bf 0})\vert^2 
({m\alpha^2\over 2n^2})\quad \\
\delta_{V_1} E^{(1)}_1=& {11m\alpha^3\over 18\pi}\Bigl( 1
-{9\pi\over 22}\xi
+{12\over 11}\xi^2 -{6\pi\over 11}\xi^3 & -{3( 2-\xi^2-4\xi^4
) \over
11\sqrt{\xi^2-1}}\tan^{-1}\sqrt{\xi^2-1}\Bigr)\\
& \xi :={2m_{e}\over m\alpha}  
 \quad & \delta_{V_1} \Gamma^{(1)}_n=0 \quad\quad\quad
\eea
where $\Psi_{n}({\bf x})$ is the Coulomb wave function.
\medskip

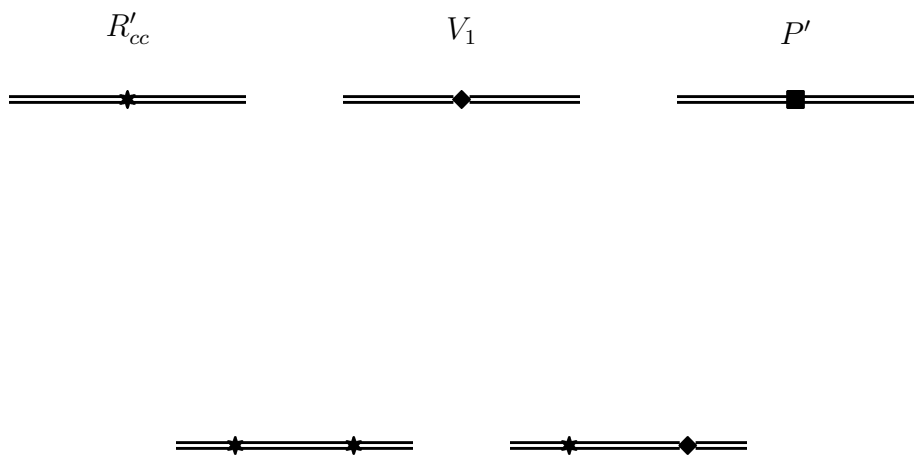
\begin{figure}
 \begin{fmffile}{2fqm}
 \begin{center}
 \begin{fmfgraph*}(90,90)
 \fmfleft{i}\fmfright{o}\fmftop{t}\fmfbottom{b}
 \fmf{dbl_plain}{i,v,o}
 \fmf{phantom}{t,v1,v2,b} 
 \fmfv{label=$R_{cc}^{\prime}$,label.side=left}{v1}
 \fmfv{decor.shape=hexagram,decor.filled=full,decor.size=50}{v}
 \end{fmfgraph*}
\hspace*{1cm}
 \begin{fmfgraph*}(90,90)
 \fmfleft{i}\fmfright{o}\fmftop{t}\fmfbottom{b}
 \fmf{dbl_plain}{i,v,o} 
 \fmf{phantom}{t,v1,v2,b} 
 \fmfv{label=$V_1$,label.side=left}{v1}
 \fmfv{decor.shape=diamond,decor.filled=full,decor.size=50}{v}
 \end{fmfgraph*}
\hspace*{1cm}
 \begin{fmfgraph*}(90,90)
 \fmfleft{i}\fmfright{o}\fmftop{t}\fmfbottom{b}
 \fmf{dbl_plain}{i,v,o} 
 \fmf{phantom}{t,v1,v2,b} 
 \fmfv{label=$P^{\prime}$,label.side=left}{v1}
 \fmfv{decor.shape=square,decor.filled=full,decor.size=50}{v}
 \end{fmfgraph*}
 \end{center}
\vspace*{0.5cm}
 \begin{center}
 \begin{fmfgraph*}(90,90)
 \fmfleft{i}\fmfright{o}\fmftop{t1,t2}\fmfbottom{b1,b2}
 \fmf{dbl_plain}{i,v1,v2,o} 
 \fmf{phantom}{t1,v3,v1,v4,b1}
 \fmf{phantom}{t2,v5,v2,v6,b2}
 \fmfv{decor.shape=hexagram,decor.filled=full,decor.size=50}{v1,v2}
 \end{fmfgraph*}
\hspace*{1cm}
 \begin{fmfgraph*}(90,90)
 \fmfleft{i}\fmfright{o}\fmftop{t1,t2}\fmfbottom{b1,b2}
 \fmf{dbl_plain}{i,v1,v2,o} 
 \fmf{phantom}{t1,v3,v1,v4,b1}
 \fmf{phantom}{t2,v5,v2,v6,b2}
 \fmfv{decor.shape=hexagram,decor.filled=full,decor.size=50}{v1}
 \fmfv{decor.shape=diamond,decor.filled=full,decor.size=50}{v2}
 \end{fmfgraph*}
 \end{center}
 \end{fmffile}
\label{fig3}
\caption{Diagrams contributing to the leading order corrections in $\Delta m/m$,
 $\alpha$ and $m\alpha^2 /\Delta m$ to the energy and decay width. The double 
line is the Coulomb propagator of the $\pi^+$ $\pi^-$ pair.}
\end{figure}

The diagrams in the second line of Fig. 3 correspond to second order perturbation theory and are not so easily calculated.
The second diagram gives a finite contribution 
$ \delta_{V_1} \Gamma^{(2)}_n$
which for the ground state
 has been evaluated numerically in \cite{Labelle}. The first
diagram has also been considered before \cite{Holstein}. However, since it is 
UV divergent a
suitable regularisation and renormalisation scheme must be specified. The subtraction
point dependence of the result will eventually cancel against the subtraction 
point
dependences in the matching coefficients. The matching coeficients are to be found
by matching the Chiral Lagrangian with e.m. interactions with (\ref{iso}) and (\ref{isob})
 which requires a
two loop calculation yet to be carried out. The matching calculation is most
efficiently done at threshold using DR and $ MS$ (or $\overline{MS}$) scheme for
 both UV and IR divergences \cite{ManoharMatch}. This requires to use the same regularisation and
 renormalisation
scheme when calculating in the effective theory. We have calculated in Appendix C the first
diagram using DR and $MS$ scheme so that our results can
be readily applied once the above mentioned matching calculation is carried out.
 We obtain

\bea
\delta_{R_{cc}^{\prime}} E^{(2)}_n=&& Re\bigl({R_{cc}^{\prime}}^2\bigr){m^2\alpha \Delta_{n}\over 4\pi}
\vert \Psi_{n}({\bf 0})\vert^2 \nn\\
 \delta_{R_{cc}^{\prime}} \Gamma^{(2)}_n=&& -2Im\bigl({R_{cc}^{\prime}}^2\bigr)
{m^2\alpha \Delta_{n}\over 4\pi}\vert \Psi_{n}({\bf 0})\vert^2
\eea 
where  $\Delta_{n}$ is given in formula (C.12) of Appendix C.

Putting all this together, our final expressions for the energy and the decay width read
\bea
E_{n}=&&-{m\alpha^2\over 4n^2}-{\vert \Psi_{n}({\bf 0})\vert^2\over 2f^2}+\delta_{V_1} E^{(1)}_{n}
\\
\Gamma_{n}=&&\Gamma_{n}^{(0)}\Biggl(1+\Delta_{\chi PT}  
+{5\Delta m\over 12m}\nn\\
&& -{m\alpha^2\over 16 \Delta m n^2}  -{m^2\alpha\Delta_{n}\over 4\pi
 f^2} \Biggr) +\delta_{V_1} \Gamma_{n}^{(2)}\\
&& \Gamma_{n}^{(0)}={9m\sqrt{2 m \Delta m }\over 64\pi f^4}\vert \Psi_{n}({\bf 0})\vert^2 
\eea
where we have substituted $R_{00}$, $R_{cc}$, $R_{0c}$ and $S_{0c}$ by their tree level values in the corrections 
\bea
 R_{00} \sim && {1\over{16f^2}}\nn\\
 R_{0c} \sim && {3\over{8f^2}}\nn\\
 R_{cc} \sim &&{1\over{2f^2}}\\
 S_{0c} \sim && -{1\over 32 m^2 f^2}\nn
\eea
and defined
\be
\vert  R_{0c}\vert^2= \left( {3\over{8f^2}} \right)^2(1 +\Delta_{\chi PT})
\ee
We have also dropped terms proportional to $R_{00}$ in (3.2) because they are  suppressed by extra factors of $m^2/4\pi f^2$. 
$\Delta_{\chi PT}$ summarizes all the contributions to  
$\vert  R_{0c}\vert^2$
beyond the one at tree level in the isospin symmetric limit, in particular
those
from pion and photon loops in the Chiral Lagrangian. $\delta_{V_1} E^{(1)}_{n}$ is given for $n=1$ in (5.3) and 
$\delta_{V_1} \Gamma_{n}^{(2)}$ is only known for $n=1\;$ \cite{Labelle}.

\be
\delta_{V_1} \Gamma_{1}^{(2)} \sim 0.4298\alpha \Gamma_{1}^{(0)}
\ee
Formulas (5.6) and (5.7) are exact up to next to leading order in $\alpha$, $\Delta m /m$ and $m\alpha^2/\Delta m$, except for the optional substitutions mentioned above.
Notice that only $R_{0c}$ is needed beyond tree level ($\Delta_{\chi PT}$). In $ \Delta_{\chi PT}$ there 
should be a contribution $\sim \log{m/\mu}$ which cancels the $\mu$ dependence in $\Delta_{n}$. This would 
arise from a two loop calculation involving photons which has not been carried out yet.

\bigskip
\section{Discussion}
\indent
\bigskip

We have presented an approach to pionium which consists of separating
the various dynamical scales involved in the problem by using effective field theory
techniques. The main advantadge of this approach is, apart from its simplicity, that
error estimates can be carried out very easily. 
A few remarks
concerning other approaches are in order. First of all, relativistic aproaches \cite{Sazdjian, Akaki}
, apart from being technically more involved, have all the scales in the problem
entangled which makes very difficult to estimate errors or to gauge the size of a
given diagram. We would like to emphasize that Lorentz symmetry, even though it is not
linearly realised, it is implemented in our approach to the required order. Several
non-relativistic approaches have appeared in the literature addressing particular aspects
of the problem \cite{Ravndal, Holstein, Labelle}. Our analysis shows that a coupled channel approach to pionium
\cite{Holstein} is unnecessary because the $\Delta m$ is much larger than the bound state energy. It
also shows that, although it is technically possible (trivial in fact) to make a
resummation of bubble diagrams {\it \`a la } Lipmann-Schwinger, it does not make much
sense doing it since there are higher derivative terms in the effective lagrangian, 
which have been neglected, which would give rise to contributions of the same order.
In a way, our approach implements the remark of \cite{Ravndal} related to the fact
that neutral pion loops give rise to important contributions in the non-relativistic
regime. We have supplemented this remark with a full theoretical framework and with
relativistic corrections of the same order which had been overlooked. 

\medskip

We have refrained ourselves to write our final results in terms of physical amplitudes because 
no experimental information is available for them at threshold energies. Simple minded identifications 
may be dangerous. For instance, one may be tempted to identify ${R_{cc}^{\prime}}$ with the strong 
scattering length. This is correct at leading order but not at the order we are interested in. 

\medskip

On the technical side we have worked out a new method to  calculate 
the Coulomb propagator $G_{c}(0,0;E)$ 
in DR. The expresions for $G_{c}(0,0;E)$ when $E\rightarrow E_{n}$ are
easily obtained for any $n$. Using DR here it is not just a matter of taste. Eventually a two
 loop matching calculation is to be done in order to extract the parameters of the Chiral
 Lagrangian from the pionium width. These kind of calculations are only efficiently done 
in DR. Since the matching coefficients depend on the renormalisation scheme, it is important
 to have our calculation in DR in order to be able to use the outcome of such a matching 
calculation straight away.

\medskip

While this paper was being written up ref. \cite{Alex} appeared which deals with the same 
problem by similar techniques.

\bigskip

{\bf Acknowledgments}

\medskip

J.S. acknowledges discussions with A. Gall, J. Gasser, V. E. Lyubovitskij and A. Rusetsky. D.E. has benefited 
from a Spanish MEC FPI fellowship.
Finantial support from the CICYT (Spain), contract AEN98-0431, and the CIRIT (Catalonia), contract 
1998SGR 00026 is also acknowledged.

\bigskip

\appendix

\Appendix{: Lorentz symmetry in non-relativistic effective theories}
\indent

\bigskip
Consider $\phi (x)$ a relativistic spin zero field and its partition function
\be
Z(J)=\int D\phi e^{i( S(\phi) +\int d^4 x J(x)\phi (x) )}
\ee
If $S$ is Lorentz invariant then
\be
Z(J)=Z(J^{\prime}) \quad\quad ,\quad\quad J^{\prime}(x)=J(\Lambda^{-1}x)
\ee
In the non-relativistic regime we only need a subset of $J$s which generate Green
functions with the external legs almost on shell. These may be choosen as
\be
J(x)=\sqrt{2m}(e^{-imx^0} J_{h}(x) +e^{imx^0} J_{h}^{\dagger}(x))
\ee
where $m$ is the mass of $\phi$ and $J_{h}(x)$ is slowly varying (i.e. contains
 energy and momentum much smaller than $m$). From (A.2) and (A.3) one easily finds
that for Lorentz transformations close to the identity
\be
J_{h}(x) \longrightarrow J_{h}^{\prime}(x)=e^{-im(\Lambda^{-1}-1)^{0}_{\; \mu}x^{\mu}} 
J_{h}(\Lambda^{-1} x)
\ee
In the non-relativistic regime $Z(J)$ can be approximated to the desired order of
accuracy by
\be
Z(J)\sim Z_{NR}(J_{h},J_{h}^{\dagger})=\int Dh dh^{\dagger}
e^{i(S_{NR}(h,h^{\dagger})+\int d^4 x (h^{\dagger}(x) J_{h}(x)+J_{h}^{\dagger}(x)h(x)))}
\ee
Then $ Z_{NR}(J_{h},J_{h}^{\dagger})$ must be invariant under the transformation
(A.4). Invariance of the terms coupled to the sources implies the following
transformations for $h(x)$.
\be
h(x) \longrightarrow h^{\prime}(x)=e^{-im(\Lambda^{-1}-1)^{0}_{\; \mu}x^{\mu}} 
h(\Lambda^{-1} x)
\ee
Hence $S_{NR}(h,h^{\dagger})$ must be constructed in such a way that it is invariant
under (A.6). In order to do so, notice first of all that $\partial_{\mu} h(x)$ does not
transform in a way similar to $h(x)$. We would like to introduce a kind of covariant
derivative. The following operator appears to be a succesful candidate
\be
D=i\partial_0 -{\partial_{\mu}\partial^{\mu}\over 2m}
\ee
We have under Lorentz transformations
\be
Dh(x)\longrightarrow e^{-im(\Lambda^{-1}-1)^{0}_{\; \mu}x^{\mu}}
 (D+i(\Lambda^{-1}-1)^{0}_{\; \mu}\partial^{\mu} ) h(\Lambda^{-1} x)
\ee
which upon the change $x\rightarrow \Lambda x$ becomes
\be
Dh(x)\longrightarrow e^{-im(1-\Lambda)^{0}_{\; \mu}x^{\mu}}
 Dh(x)
\ee
Analogously, if we have  $C_{w}(x)=(h^{\dagger}(x))^{m}(h(x))^n$ , $w=n-m$, we
may define for $w\not=0$
a generalisation of (A.7 )
\be
D=i\partial_0 -{\partial_{\mu}\partial^{\mu}\over 2wm}
\ee
Then $D^{k} C_{w}(x)$ has the same transformation properties as $C_{w}(x)$. We call
$w$ the weight of the composite field $C_{w}(x)$. If $w=0$ then $\partial_{\mu} C_0(x)$
transforms as a usual Lorentz vector.

From the discussion above the following rules can be inferred in order to built a
Lorentz invariant non-relativistic effective theory for spin zero particles:

(i) Write down all possible terms in the particle sector we are interested in with weight
 zero and no derivatives up to the desired order.

(ii) For each term, which is not already of the higher relevant order, insert $D$s or
$\partial_{\mu}$s in all possible ways. All $\mu$ indices coming from the $\partial_{\mu}$  
must be contracted in a Lorentz invariant way.

Applying the rules above we obtain the lagrangians (2.3) and (2.4). Recall also that for 
the particular case we are interested in the (minimal)
supression of $D$ is $\Delta m/m$ whereas the (minimal) suppresion of $\partial_{\mu}$ is 
$\sqrt{\Delta m/m}$.

Finally, let us mention that for practical purposes the rules that we have obtained are
identical to those derived from the so called reparametrisation invariance \cite{LM} (see also \cite{RI}).
 Hence,
it should be clear that reparametrisation invariance is nothing but a way to implement
Lorentz symmetry in a non-relativistic theory. We believe that this point is important and
has not been sufficiently stressed in the literature.

\bigskip
\Appendix{: Local field redefinitions}
\indent

\bigskip 

The lagrangian given in formulas (\ref{iso}) and (\ref{isob}) contains higher 
time derivative terms whereas the
usual non-relativistic lagrangians contain only a time derivative in the bilinear terms of 
each field. The latter is known as the minimal form of the lagrangian. In this Appendix we
display the local field redefinitions which bring the lagrangian (\ref{iso}) and
 (\ref{isob}) to its minimal form.
Let us only mention that local field redefinitions exploit the freedom we have in field theory to 
choose the interpolating field we wish, and refer the interested reader to the literature 
on the subject \cite{LM,Balzereit,Georgi,FR,Ecker}. 
The price we pay for having the lagrangian in its minimal form is that Lorentz symmetry 
(reparametrization invariance) will not be explicit anymore. The constraints given by Lorentz symmetry
 will reduce to non-trivial relations between the parameters of the lagrangian in its minimal form.

\medskip

We are retaining corrections up to the relative order 
$(\Delta m /m)^{\frac {3}{2}}$.
In order to
 reduce our operator basis, we will take 
advantadge of the fact that local field
redefinitions can also be organised in powers of $\Delta m /m$. The induced terms beyond the desired order 
as well as the terms which do not contribute to the two particle sector (six pion terms and beyond) 
will be neglected.

\medskip

Let us first consider local field redefinitions which keep Lorentz symmetry explicit. We can get rid of 
the $A_0$ and $\delta_2$ terms in  (\ref{iso}) and (\ref{isob}) by 
 
 \bea
 \bfpi^{i}& \mapsto &((1-{D A_0\over 2})\d^{ij} \bfpi^{j} +({\d_1 A_0 {\bf Q}^{i} {\bf Q}^{j}\over2} - 
 {\d_2 {\bf Q}^{i} {\bf Q}^{j}\over 2}))\bfpi^{j}
 \eea
The  bilinear terms become
 \bea
{\rm L}_2 +\Delta{\rm L}_2  = &&\bfpi^{\dagger} D \bfpi + {\bfpi^{\dagger}}^{i} \d_1 {\bf Q}^{i}{\bf Q}^{j} (1 +
 (\d_1 A_0 - \d_2){\bf Q}^2) {\bf \pi^{j}}
 \eea
 and the following constants of the four pion terms get modified 

\bea
A_1 \rightarrow && A_1^{\prime}=A_1 - A_0 B_2\nn\\
A_2 \rightarrow && A_2^{\prime}=A_2 - A_0 B_1\nn\\
C_1 \rightarrow && C_1^{\prime\prime}=C_1 -(\delta_2-\delta_1 A_0)B_2 \\
C_2 \rightarrow &&  C_2^{\prime\prime}=C_2 -2(\delta_2-\delta_1 A_0)B_1 \nn
\eea

We can also get rid of the $A_1^{\prime}$ and $A_2^{\prime}$ keeping Lorentz invariance by making

\bea
 \bfpi \mapsto \bfpi  - {A_2^{\prime}}^{\ast} 
\bfpi (\bfpi^
 {\dagger}\bfpi)  - {A_1^{\prime}}^{\ast} 
\bfpi^{\dagger} (\bfpi\bfpi) 
 \eea
which induces
\bea 
C_1^{\prime\prime} \rightarrow && C_1^{\prime\prime\prime}=C_1^{\prime\prime} - A_1^{\prime} \delta_1\nn\\
C_2^{\prime\prime} \rightarrow &&  C_2^{\prime\prime\prime}=C_2^{\prime\prime} - (A_2^{\prime}+ {A_2^{\prime}}^{\ast})
\delta_1 
\eea 

The remaining time derivatives in $D$ and in the $A_3$ and $A_4$ terms can only be removed if we give up the
explicit realisation of Lorentz symmetry which we have kept so far. Notice that the time derivatives in the $A_4$
term are higher order and can be dropped. The following field redefinition gets rid of the higher order time derivatives
in the bilinear terms 
\bea
 \bfpi^{i}& \mapsto &\Bigl((1 - {i\partial_0\over 4m} + {\bfnabla^2\over 8m^2})\delta^{ij}+{\d_1 
{\bf Q}^{i} {\bf Q}^{j}\over 4m}
\Bigr)\bfpi^{j}
 \eea
Finally the time derivatives induced by this redefinition in the four pion terms together with the remaining time derivatives in
$A_3$ and $A_4$ can be removed by

 \bea
  \bfpi \mapsto \bfpi + ({B_1\over 2m}  - A_5 )\bfpi (\bfpi^
 {\dagger}\bfpi) + ({B_2\over 2m} - A_3 )\bfpi^{\dagger} (\bfpi{\bf
 \bfpi})
\eea

Putting all together, we obtain for the constants in the lagrangian (\ref{minimal}) the following expressions in terms of
the original constants

 \bea
\Delta m =&& \d_1 {\bf Q}^2 \left(1+\left(\d_1 A_0 -\d_2 +{\delta_1\over 2m}\right){\bf Q}^2 \right)\nn\\
 D_1 =&& {B_1\over m}- A_5+2mA_4 \nn\\
 D_2 =&& {B_2\over m}- A_3 \\
 C_1^\prime =&& C_1-B_2\d_2+({B_2  \over m}  - A_3   - A_1  +2 A_0 B_2) \d_1\nn\\
 C_2^\prime =&& C_2-2B_1\d_2+({2 B_1 \over m} - A_2 -A_2^{\ast} + 4 A_0 B_1  - 2 A_5)\d_1 \nn
 \eea

Upon restricting the lagrangian (\ref{minimal}) to the zero charge sector we obtain the lagrangian (\ref{zero})
the constants of which are related to the above ones according to  
\bea
 R_{00} =&& B_1 + B_2 + e^{2}(C_1^{\prime}+{C_1^{\prime}}^{\ast}) +e^{2}C_2^{\prime}\nn\\
 R_{0c} =&& 2 B_2+2e^2{C_1^{\prime}}^{\ast}\nn\\
 R_{cc} =&& 2 B_1 + 4 B_2 + 2e^{2}C_3 \nn\\
 S_{00} =&&  {D_1\over {2m}} + {D_2\over {2m}}\\
 S_{0c} =&& {D_2\over 2m}\nn\\
 S_{cc} =&& {D_1\over 2m} +{D_2\over m} \nn\\
 P_{00} =&&  2 A_4 \nn\\
 P_{cc} =&&  2 A_4 \nn
 \eea

\bigskip

\Appendix{: The Coulomb propagator in D space dimensions}
\indent

\bigskip
We present here a generalisation of the Coulomb propagator to $D$ space dimension which may prove useful in
bound state calculations. For the actual Coulomb potential in $D$ dimensions
\be
V_c(r)=-{\alpha c_{D}\over r^{D-2}} \quad\quad ; \quad\quad  c_{D}={4\pi\Gamma ({D\over 2})\over (D-2)2\pi^{D\over 2}}
\ee
we have not been able to find an explicit representation. However, a slight modification of it
\be
V_c (r)\rightarrow V_c^{\prime}(r)=-{\alpha  c^{\prime}_{D}\over r} \quad\quad ; \quad\quad 
 c^{\prime}_{D}={4\pi\over \Gamma ({D-1\over 2}) (4\pi)^{D-1\over 2}}
\ee
admits the following exact representation, which is a generalisation of that presented in
\cite{Voloshin}
\be
G_{c}({\bf x},{\bf y},E)=\sum_{l=0}^{\infty}G_{l}( x, y,E)\sum_{\{m_{i}\}}
Y_{l}^{\{m_{i}\}}({{\bf x}\over x})
{Y_{l}^{\ast}}^{\{m_{i}\}}({{\bf y}\over y}) 
\ee
\be
G_{l}( x, y,E)= -m(2k)^{D-2}(2kx)^{l}(2ky)^{l} e^{-k(x+y)}\sum_{s=0}^{\infty}{L_{s}^{2l+D-2}(2kx)
L_{s}^{2l+D-2}(2ky) \Gamma (s+1) \over (s+{2l+D-1\over 2}-{m\alpha\over 2k}c^{\prime}_{D} )
\Gamma (s+2l+D-1) } \nn
\ee
where $Y_{l}^{\{m_{i}\}}$ are the spherical harmonics in $D$ dimensions and 
$E=-k^2/m$.
The potential $ V_c^{\prime}(r)$ corresponds to the following modification of the longitudinal photon
propagator in standard DR
\be
{1\over {\bf k}^2 }\longrightarrow ({1\over 
{\bf k}^2})^{D-1\over 2}
\ee 
The change of regularisation scheme necessary for translating the result of (C.3) to those of
standard DR can be obtained by calculating the logarithmically divergent
diagram of Fig. 4  with the two propagators in (C.5). Using the $MS$ renormalisation scheme for both 
regularisations we obtain
\be
\log{\mu^{\prime}\over \mu}= {\gamma_{E}-1-\log(4\pi )\over 2}
\ee

 \begin{figure}

 \begin{fmffile}{12frs}
 \begin{center}{
 \begin{fmfgraph*}(90,90)
 \fmfleft{i1,i2}\fmfright{o1,o2}
\fmftop{u1}\fmfbottom{d1}
 \fmf{fermion,label=$\pi^{+}$,label.side=left}{i1,v1}\fmf{fermion,label=$\pi^{-}$,label.side=right}{i2,v1} 
 \fmf{fermion,label=$\pi^{+}$,label.side=left}{v2,o1}\fmf{fermion,label=$\pi^{-}$,label.side=right}{v2,o2} 
\fmf{fermion,left,width=thin
}{v1,v3,v2}
 \fmf{fermion,right,width=thin
}{v1,v4,v2}
\fmf{phantom}{u1,v3}\fmf{phantom}{d1,v4}
\fmf{dashes}{v3,v4}
\fmfv{label=$\pi^{+}$,label.dist=0.5cm}{v4}
\fmfv{label=$\pi^{-}$,label.dist=0.5cm}{v3}
\end{fmfgraph*}
 }\end{center}
 \end{fmffile}
\label{fig4}
\caption{Logarithmically divergent diagram which is calculated with the two longitudinal photon propagators (C.5) for the dashed lines.}
\end{figure}
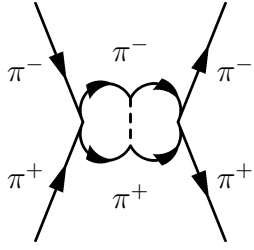

The calculation of $G_c(0,0;E)$ can be easily done using the formula 1.4.(1) of ref. \cite{tables}
\be
\sum_{n=-\infty}^{\infty} {\Gamma (a+n)\Gamma (b+n)\over 
\Gamma (c+n)\Gamma (d+n)}={ \pi^2\Gamma (c+d-a-b-1)\over \sin (\pi a)\sin (\pi b)\Gamma (c-a)\Gamma (d-a) 
\Gamma (c-b)\Gamma (d-b)}
\ee
We obtain ($D=3+2\epsilon^{\prime}$)
\be
(\mu^{\prime})^{-2\epsilon^{\prime}}G_{c}({\bf 0},{\bf 0},-{k^2\over m}) = -2mk{2\pi^{D\over 2}\over 
\Gamma ({D\over 2})}
 ({2k\over \mu^{\prime}})^{2\epsilon^{\prime}}
\sum_{s=0}^{\infty}
{\Gamma (s+D-1) \Gamma (s+{D-1\over 2}-{m\alpha\over 2k}c^{\prime}_{D} )
\over  \Gamma (s+1) \Gamma (s+{D+1\over 2}-{m\alpha\over 2k}c^{\prime}_{D} )
\Gamma^2 (D-1) }\nn
\ee
\bea
=&& {mk\over 4\pi}\Biggl( 1 +\\
&& +{m\alpha\over 2k}\Bigl({1\over \epsilon^{\prime}} +2\log({2k\over \mu^{\prime}}) 
+2\gamma_{E}-2\log(4\pi ) -2\Bigr) +\\
&& +{m\alpha\over k}\Bigl(\psi(1+{m\alpha\over 2k})-\psi (1)
+{\pi \cos({m\alpha\pi\over 2k})\over \sin  ({m\alpha\pi\over 2k})}-{2k\over m\alpha} \Bigr)\Biggr)
\eea
 (C.9), (C.10) and (C.11) correspond to zero, one and more than one longitudinal photon exchange respectively.
For $E\rightarrow E_n=m\alpha^2/4n^2$ we have
\bea
 \lim_{E\rightarrow E_n} \Bigl( (\mu^{\prime})^{-2\epsilon^{\prime}}
G_{c}({\bf 0},{\bf 0},-{k^2\over m})-{\Psi_{n}
({\bf 0})\Psi_{n}^{\ast} ({\bf 0})\over E-E_{n}}\Biggr) =&& {m^2\alpha\over 8\pi}\Bigl( {1\over n} +\nn\\
&& + \bigl( 
2\log({m\alpha\over n \mu}) 
+\gamma_{E}-\log(4\pi ) -1 \bigr)+\nn\\
&& + \bigl( 2\psi (n)+
2\gamma_{E}-{3\over n} \bigr)\Bigr)\\
=:&& {m^2\alpha \Delta_{n}\over 4\pi}\nn
\eea
where we have used the MS renormalisation scheme and changed $\mu^{\prime}$ by $\mu$ according to (C.6) so 
that the results above are in standard DR with MS scheme.  Clearly the
singular part is local, independent of the principal quantum number $n$, and can be absorbed in a renormalisation 
of $R^{\prime}_{cc}$.
 This result is in agreement with a recent DR calculation
of the same object carried out in \cite{ma6}. Finally formula (5.5) is obtained.




\begin{thebibliography}{99}

\bibitem{Deser} S.Deser et al., {\it Phys.Rev.} {\bf 96} (1954) 774;
 T.L.Trueman, {\it Nucl.Phys.} {\bf 26} (1961) 57;
 E.Lambert, {\it Helv.Phys.Acta} {\bf 42} (1969) 667;
 H.Pilkuhn and S.Wycech, {\it Phys.Lett.} {\bf 76B} (1978) 29;
 G.Rasche and W.S.Woolcock, {\it Nucl.Phys.} {\bf A381} (1982) 405.

\bibitem{Rasche}  A. Gashi, G. Rasche, G. C. Oades and  W. S. Woolcock, {\it Nucl.Phys.}{\bf A628} (1998) 101;
G. Rasche and A. Gashi, {\it Phys.Lett.}{\bf B404} (1997) 375.

\bibitem{Dirac} B. Adeva et al.,
{\it CERN-SPSLC-}{\bf 95}-{\bf 1}
; J. Schacher, hep-ph/9808407.

\bibitem{Nemenov} L.L. Nemenov, {\it Yad. Fiz.} {\bf 41} (1985) 980.

\bibitem{GL} J.Gasser and H.Leutwyler, {\it Ann. of Phys.} {\bf 158} (1984) 142.

\bibitem{Lepage} W.E. Caswell and G.P. Lepage, {\it Phys. Lett.} {\bf
B167} (1986) 437.


\bibitem{pionium} L. Afanasyev et al., {\it Phys. Lett.} {\bf B 308} (1993) 200; {\bf B 338} (1994) 478. 

\bibitem{Sazdjian} H. Jallouli and H. Sazdjian, {\it Phys.Rev.} {\bf D58} (1998) 014011; Erratum-ibid. 
{\bf D58} (1998) 099901.

\bibitem{Akaki}  M. A. Ivanov, V. E. Lyubovitskij,
 E. Z. Lipartia and A. G. Rusetsky,
{\it Phys.Rev.}{\bf D58} (1998) 094024.

\bibitem{NRQED} T. Kinoshita and M. Nio,
{\it Phys. Rev.} {\bf D53} (1996) 4909; P. Labelle, S.M. Zebarjad and C.P. Burgess,
{\it Phys. Rev.} {\bf D56} (1997) 8053; A.H. Hoang, P. Labelle and S.M. Zebarjad, 
{\it Phys.Rev.Lett.} {\bf 79} (1997) 3387.

\bibitem{ManoharMatch}  A.V. Manohar, {\it Phys. Rev.} {\bf D56} (1997) 230; A. Pineda and J. Soto,
{\it Phys.Rev.} {\bf D58} (1998) 114011.


\bibitem{LM} M. Luke and A. V. Manohar, {\it Phys.Lett.} {\bf B286} (1992) 348.

\bibitem{isob}  H. B. O'Connell, K. Maltman, A. W. Thomas and A. G. Williams, 
hep-ph/9707404. 

\bibitem{2loop} J. Bijnens, G. Colangelo, G. Ecker, J. Gasser and M. E. Sainio,
 {\it Nucl.Phys.} {\bf B508} (1997) 263; Erratum-ibid. {\bf B517} (1998) 639;
{\it Phys.Lett.} {\bf B374} (1996) 210. 

\bibitem{Knech} M. Knecht and R. Urech, {\it Nucl.Phys.} {\bf B519} (1998) 329.

\bibitem{Ravndal}  X. Kong and F. Ravndal, hep-ph/9805357.

\bibitem{Holstein}  B. R. Holstein, nucl-th/9901041.

\bibitem{LambPos}  A. Pineda and J. Soto,
{\it Phys. Lett.} {\bf B420} (1998) 391; {\it Phys.Rev.} {\bf D59} (1999) 016005.

\bibitem{Labelle}  P. Labelle and K. Buckley, hep-ph/9804201.

\bibitem{Alex} A. Gall, J. Gasser, V. E. Lyubovitskij and A. Rusetsky, hep-ph/9905309.

\bibitem{Balzereit} Ch. Balzereit, hep-ph/9809226; W. Kilian and T. Ohl, Phys. Rev. D50 (1994) 4649.
 C. Balzereit and T. Ohl, 
{\it Phys. Lett.} {\bf B386} (1996) 335.


\bibitem{RI} R. Sundrum {\it Phys.Rev.} {\bf D57} (1998) 331; M. Finkemeier, H. Georgi and M. McIrvin 
{\it Phys.Rev.}{\bf D55} (1997) 6933; C. L. Lee, hep-ph/9709238.

\bibitem{Georgi} H. Georgi, {\it Nucl. Phys.} {\bf B361}, 339 (1991).

\bibitem{FR}  S. Scherer and H.W. Fearing, 
{\it Phys. Rev.} {\bf D52} (1995) 6445.

\bibitem{Ecker}  G. Ecker and M. Mojzis {\it Phys.Lett.} {\bf B410} (1997) 266, 
ibid. {\bf B438} (1998) 446 (Err.).

\bibitem{Voloshin}  M. B. Voloshin,
{\it Sov. J. Nucl. Phys.} {\bf36} (1982) 143; A. Pineda, {\it Nucl.Phys.} {\bf B494} (1997) 213.

\bibitem{tables} H. Bateman, {\it Higher Transcendental Functions}, Vol. I, McGraw-Hill, 1953.

\bibitem{ma6} A. Czarnecki, K. Melnikov, A. Yelkhovsky, hep-ph/9901394.


\end{thebibliography}
\end{document}